# Localization of light in an optical microcapillary introduced by a droplet

Tabassom Hamidfar[1,2], Kirill V. Tokmakov[1], Brian J. Mangan[3], Robert S. Windeler[3],
Artemiy V. Dmitriev[1,4], Dashiell L. P. Vitullo[1], Pablo Bianucci[2], and Michael Sumetsky[1,*]

[1]Aston Institute of Photonic Technologies, Aston University, Birmingham B4 7ET, UK
[2]Concordia University, Montreal, Québec H4B 1R6, Canada
[3]OFS Laboratories, 19 Schoolhouse Road, Somerset, NJ 08873, USA
[4]School of Physics and Astronomy, University of Birmingham, Birmingham B15 2TT, UK
*Corresponding author: m.sumetsky@aston.ac.uk

**Abstract**

Sensing with optical whispering gallery modes (WGMs) is a rapidly developing detection method in modern microfluidics research. This method explores the perturbations of spectra of WGMs propagating along the wall of an optical microcapillary to characterize the liquid medium inside it. Here we show that WGMs in a silica microcapillary can be fully localized (rather than perturbed) by evanescent coupling to a water droplet and, thus, form a high quality-factor microresonator. The spectra of this resonator, measured with a microfiber translated along the capillary, present a hierarchy of resonances which allow us to determine the size of the droplet and variation of its length due to the evaporation. The resolution of our measurements of this variation equal to 4.5 nm is only limited by the resolution of the optical spectrum analyzer used. The discovered phenomenon of complete localization of light in liquid-filled optical microcapillaries suggests a new type of microfluidic photonic devices as well as an ultraprecise method for microfluidic characterization.

## 1. Introduction

Research and development of optical whispering gallery mode (WGM) microresonators has grown significantly over the last several decades finding applications in telecommunications[1-4], fabrication of microlasers[5-8], quantum electrodynamics[9, 10], quantum networks[11, 12] microfluidics [13-16], optomechanics[17] and other fields of science and engineering. These microresonators are usually fabricated in the form of spheres, toroids, bottles, bubbles, cylinders, disks, and capillaries (see e.g., Refs [1, 9, 13, 18]).

Similar to the acoustic WGMs discovered by Lord Rayleigh more than 100 years ago[19], optical WGMs propagate very close to the surface of optical microresonators. This fact has both positive and negative consequences for applications. On the one hand, since WGMs are isolated from the microresonator core, thermal and electrical tuning devices can be included into interior part of the microresonator without degrading its quality factor[20,21]. Of special interest is the development of microfluidic optomechanics based on liquid-filled microcapillaries which is also based on the effect of isolation of WGMs from the liquid inside the resonator[22].

Alternatively, coupling between WGMs and the interior of the microresonator is critical for the development of ultraprecise capillary microfluidic sensors[13, 16, 23, 24]. To this end, researchers fabricate optical microcapillaries having very narrow (a few micron thick) walls which allow WGMs to evanescently penetrate through the internal capillary surface and probe the interior medium. It is expected that these microfluidic sensors, similar to WGM microresonators sensing the environment at their peripheral surface[15], can achieve breakthrough single molecule sensitivity.

Light propagating in the form of a WGM along an optical fiber surface can be fully controlled by the exceptionally small nanoscale variation of the effective fiber radius[25]. In particular, the deformation of a fiber by just a fraction of a nanometer can completely localize WGMs and form a high quality-factor microresonator[26]. This means that the fiber radius variation sufficient for governing WGMs can be dramatically smaller than the wavelength of light. The technological platform based on this phenomenon – Surface Nanoscale Axial Photonics (SNAP) – demonstrates the fabrication of miniature resonant optical devices at the fiber surface with unprecedented sub-angstrom precision and ultralow loss[3, 25, 27].

The success of the work presented below is primarily based on the idea of bridging the SNAP platform[25] and WGM microfluidic sensing platform[23], which have much in common[28]. In fact, while SNAP considers WGMs controlled by nanoscale deformation of the *outer surface* of an optical fiber, similar control can be performed in microcapillaries by small and slowly varying *internal nonuniformities*. Recently we showed that the introduction of SNAP resonators is possible by deformation of the external as well as the internal capillary surfaces[29]. Alternatively, here we consider liquid situated in a *uniform* microcapillary. We show that the control over the WGMs propagating along the microcapillary can be performed only by the nonuniformities of the liquid. In particular, we demonstrate that a water droplet situated inside a silica microcapillary can fully localize WGMs and introduce a high quality-factor optical microresonator. Our discovery paves the way for the development of a novel comprehensive method for microfluidic sensing and a new type of tunable resonant microfluidic-based photonic devices.

## 2. Observation of optical microresonator introduced by a droplet

The microcapillary used in our experiment was prepared from a polymer-coated silica capillary fiber drawn at OFS Laboratories. Fig. 1a shows the optical microscope image of the cross-section of this capillary after the polymer coating was removed in hot sulfuric acid. From this image, the external and internal radii of the capillary were determined as $r_{int} = 61\,\mu\text{m}$ and $r_{ext} = 68\,\mu\text{m}$. In our experiment, one end of a several centimeters long fiber was glued in into the needle of a syringe that was used to launch and position the water droplets inside the fiber. The other end was left open. The silica microcapillary was prepared by removing the coating from a few centimeter section in the middle of this fiber. An optical microscope image of a sample droplet positioned inside the stripped section of the fiber is shown in Fig. 1b. It is seen that the droplet possesses the characteristic meniscuses introduced by the surface tension.

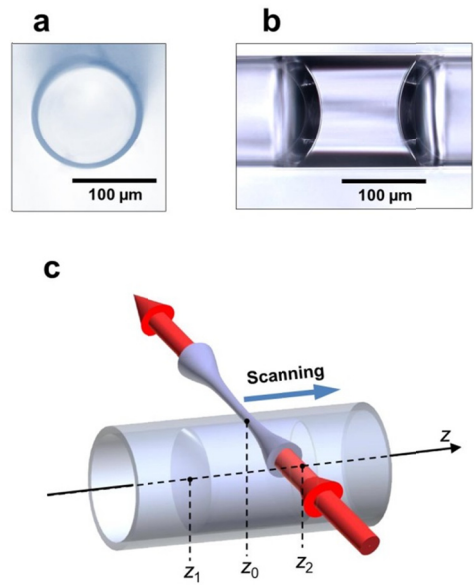

**Fig. 1. a** The optical microscope image of the cross-section of the silica microcapillary used in the experiment. This image was used to determine the radius of the capillary and the thickness of its wall. **b** The optical microscope image of a water droplet inside the capillary. The droplet meniscuses caused by the surface tension are clearly seen. **c** Illustration of the experiment. The input-output taper with a micron-diameter waist (microfiber) is positioned normally to the capillary. The microfiber excites WGMs which propagate along the microcapillary wall and sense the droplet.

The idea of our experiment is illustrated in Fig. 1c. A biconical fiber taper with a micron-diameter waist (microfiber) was oriented normal to the capillary and connected to the LUNA optical spectrum analyzer (1.3 pm wavelength resolution). The microfiber was translated along the capillary axis $z$ and periodically placed in contact with the capillary to measure the transmission power $P(z,\lambda)$ as a function of the microfiber coordinate along the capillary $z$ and wavelength $\lambda$ [25]. Fig. 2a shows the surface plot of transmission spectra $P(z,\lambda)$ collected in the bandwidth $1539.5\,\text{nm} < \lambda < 1545\,\text{nm}$ at points spaced by 2 μm along the 600 μm capillary section with the droplet in the middle. The spectra are found to be quasiperiodic in wavelength with the period 3.92 nm (see Supplementary note 1). The time of each measurement (including the time required for recording the spectrum and the time of translation of the microfiber between the measurement points) was 10 sec. Consequently, the full time of recording of the spectra along the 600 μm capillary section was 50 min, while the measurements along the length of the droplet took around 11 min.

The spectral plot in Fig. 2a clearly outlines the area where the water droplet is situated. As explained in the next section, the green pathways (dips) outside the droplet area correspond to the cutoff wavelengths of WGMs. Some of these pathways (e.g. those close to 1540.2 nm, 1540.9 nm, and 1541.8 nm) cross the droplet area unaffected. This means that the corresponding WGMs are not perturbed by the presence of the droplet. Other lines (e.g., those close to 1539.9 nm, 1540.5 nm, and 1543.3 nm) are interrupted in the droplet area indicating that the corresponding WGMs are perturbed by the droplet. The interruption of these lines is accompanied by the appearance of additional resonances, which is explained by the full localization of the corresponding WGMs in the droplet area. The evanescent coupling of these WGMs to the water droplet causes the positive shifts of their cutoff wavelengths and the formation of a high quality-factor microresonator. The position of resonances is in excellent agreement with the theory presented below.

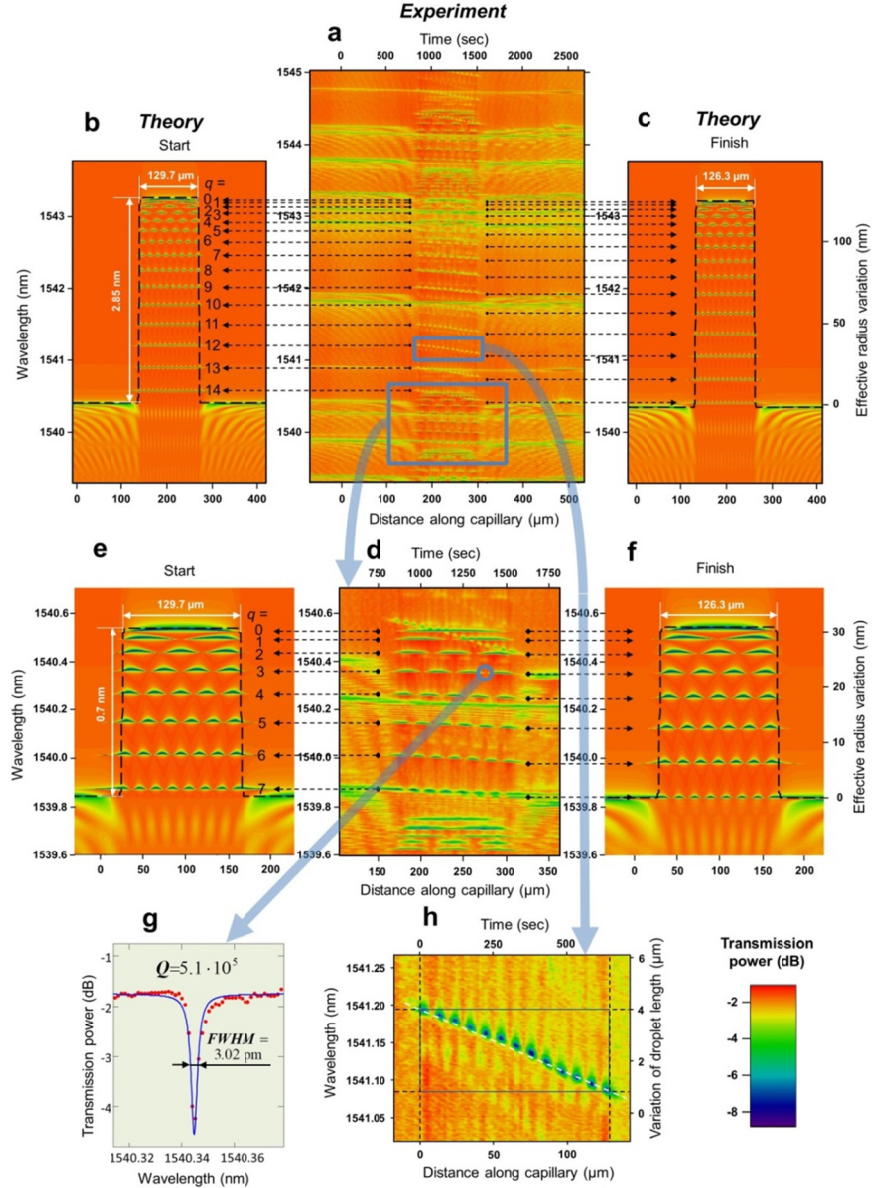

**Fig. 2.** **a** The surface plot of the transmission power spectra collected by the input-output microfiber translated along the microcapillary with a droplet inside. The spectral (vertical axis) and spatial (horizontal axis) resolutions of the scan were 1.3 pm and 2 μm, respectively. The central part of the plot indicates the region where the droplet was situated. **b, c** Surface plots of the calculated transmission power spectra corresponding to the stationary droplet with the initial (b) and final (c) dimensions which were obtained for a single cutoff wavelength using Eq. (1)-(3). The initial and final dimensions were taken from the experimental plot **a** at axial coordinates $z_1$ and $z_2$ (Fig. 1c) when the measurement of the droplet region started and finished. **d, e, f** Magnified spectral region of the surface plot **a** and the theoretical models of the droplet similar to those shown in **b** and **c** but for a different cutoff wavelength. **g** The resonance of the droplet-introduced microresonator measured at the node indicated at the spectral plot **d**. **h** Magnified spectral region of the surface plot **a** containing the pathway of resonances with the axial quantum number $q=12$ which was used for the analysis of the droplet evaporation.

## 3. Theory of cutoff wavelengths and quantization of slow WGMs

High quality-factor optical resonance lines are commonly observed in the WGM spectra of optical fibers[13, 23, 25]. These resonances correspond to the so called cutoff wavelengths of light $\lambda_{mp}$. Here $m$ and $p$ are the azimuthal and radial quantum numbers which numerate modes propagating in an axially symmetric and uniform fiber (in particular, a capillary fiber). The expression for the electromagnetic field of these modes is factorized in the cylindrical coordinates $(z, \rho, \varphi)$ as $E_{mp}^{\pm}(z, \rho, \varphi) = \exp(\pm i \beta_{mp}(\lambda) z) \exp(im\varphi) Q_{mp}(\rho)$ [30]. Here $\beta_{mp}(\lambda)$ is the propagation constant and the function $Q_{mp}(\rho)$ exponentially decreases outside the capillary for $\rho > r_{ext}$ and inside the capillary for $\rho < r_{int}$. The optical modes of our interest are the WGMs which are adjacent to the optical fiber surface. At cutoff wavelengths $\lambda_{mp}$, the propagation constants of these modes approach zero, $\beta_{mp}(\lambda_{mp}) = 0$. The speed of propagation of WGMs with small $\beta_{mp}(\lambda)$ along the capillary axis is slow. It is due to the slowness of WGMs they can be governed by angstrom-scale variations of the effective fiber radius employed in SNAP[25]. As explained below, a similar effective radius variation effect can be introduced by the shifts of cutoff wavelengths $\lambda_{mp}$ caused by the evanescent coupling of WGMs to a water droplet inside a capillary fiber.

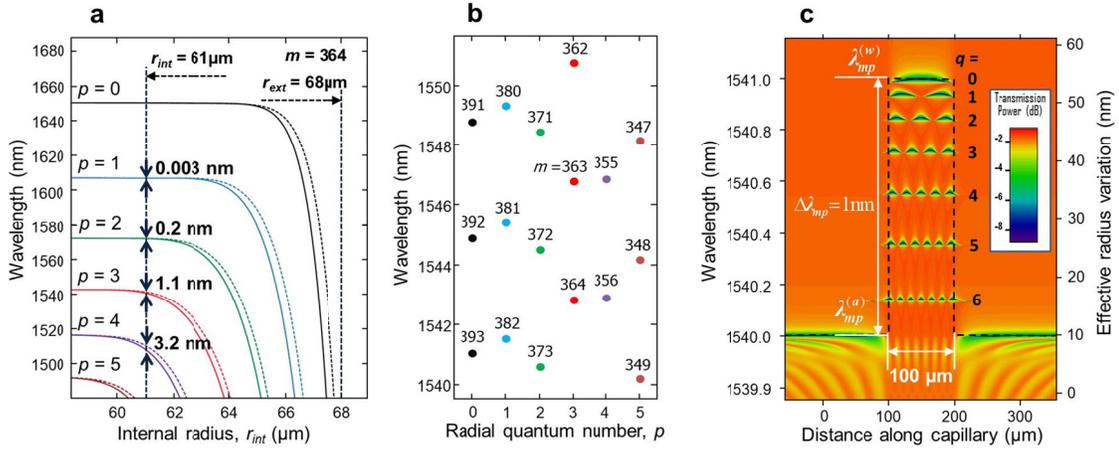

**Fig. 3.** **a** Cutoff wavelengths with the azimuthal quantum number $m = 364$ and radial quantum numbers $p = 0, 1, 2, 3, 4, 5$ for the silica microcapillary with external radius $r_{ext} = 68$ μm as a function of the microcapillary internal radius $r_{int}$. The solid and dashed curves correspond to the empty and water-filled microcapillary, respectively. **b** The distribution of cutoff wavelengths (dots) for the TE-polarized WGMs having $p = 0, 1, 2, 3, 4, 5$ in the bandwidth 1540 nm < $\lambda$ < 1550 nm. The corresponding azimuthal quantum numbers $m$ are shown next to each of the dots. **c** A sample surface plot of transmission power spectra calculated using Eqs. (1)-(3) for the water-introduced shift of the cutoff wavelength equal to 1 nm and microfiber-capillary coupling parameters indicated in the text.

Fig. 3a shows the dependencies of $\lambda_{mp}$ for the silica microcapillary with the refractive index $n_{cap} = 1.46$ and external radius $r_{ext} = 68$ μm as a function of internal radius $r_{int}$. The solid and dashed curves correspond to the empty and water-filled capillary, respectively. The external and internal radii, $r_{int} = 61$ μm and $r_{ext} = 68$ μm, of the capillary used in our experiment are indicated by the vertical dashed lines. In calculations, we chose the azimuthal quantum number $m = 364$ so that the cutoff wavelength of the first WGM sensitive to water corresponds to the wavelength ~1.54 μm close to the value of an experimental cutoff wavelength in Fig. 2a. The noticeable shifts caused by the presence of water are found for the WGMs with radial quantum numbers $p \geq 2$. Specifically, for $p = 2, 3, 4$ and 5 these shifts are 0.003 nm, 0.2 nm, 1.1 nm, 3.2 nm, and 6.0 nm, respectively.

Fig. 3b shows the calculated cutoff wavelengths $\lambda_{mp}$ for the TE polarized WGMs with radial quantum numbers $p = 0, 1, ..., 5$ which are situated in the bandwidth 1.54 μm < $\lambda$ < 1.55 μm (see Supplementary Note 2).

As we found above, the WGMs having frequencies $\lambda_{mp}$ with small $p = 0$ and 1 are not affected by liquid. On the other hand, the WGMs with large $p \gg 1$, which strongly scatter from interfaces, may be undetectable due to their high attenuation. For this reason, the values $\lambda_{mp}$ situated in the bandwidth $1.54\,\mu\text{m} < \lambda < 1.55\,\mu\text{m}$ is shown in Fig. 3b for quantum numbers $p \leq 5$, which are presumably detected in our experiment. The corresponding azimuthal quantum numbers $m$ ranging from 347 to 393 are shown in Fig. 3b as well. The cutoff wavelengths of the TM polarized WGMs, which are not shown in this figure, are known to have significantly smaller quality-factors than those of the TE polarized WGMs due to scattering from the surface and bulk nonuniformities[31]. The distribution of the cutoff wavelengths in Fig. 3b is in a reasonable quantitative agreement with that of the measured resonant wavelengths in Fig. 2a.

The appearance of a microresonator introduced by a water droplet is explained as follows. In the vicinity of cutoff wavelengths, the propagation constant is small and can be determined in the first order of the perturbation theory as (see Supplementary Note 2)

$$\beta_{mp}^{(a),(w)}(\lambda) = 2^{3/2}\pi n_{cap}\left(\lambda_{mp}^{(a),(w)}\right)^{-3/2}\left(\lambda_{mp}^{(a),(w)} - \lambda\right)^{1/2}, \tag{1}$$

where $n_{cap}$ is the refractive index of the capillary material and superscript indices $^{(a)}$ and $^{(w)}$ correspond to the air-filled and water-filled capillaries, respectively. From Eq. (1), the propagation constant is real for wavelengths $\lambda < \lambda_{mp}^{(a),(w)}$ and pure imaginary for $\lambda > \lambda_{mp}^{(a),(w)}$. Assume that the cutoff wavelength $\lambda_{mp}^{(a)}$ of the empty capillary experiences a relatively small positive perturbation and becomes equal to $\lambda_{mp}^{(w)}$ after the capillary is filled in with water. Then, for the wavelength $\lambda$ which is close to $\lambda_{mp}^{(a)}$ and $\lambda_{mp}^{(w)}$, the expressions for propagation constants in the water-filled and empty capillary can be calculated from Eq. (1). In our approximation, the relatively small perturbation of the WGM dependence on the radial coordinate in $Q_{mp}(\rho)$ due to the presence of water can be neglected. Then, the behavior of WGMs along the microcapillary with a droplet inside is determined by the one-dimensional wave equation:

$$\frac{d^2\Psi_{mp}}{dz^2} + \beta_{mp}^2(\lambda,z)\Psi_{mp} = 0 \tag{2}$$

where $\beta_{mp}(z,\lambda) = \beta_{mp}^{(w)}(\lambda)$ inside the droplet region, $z_1 < z < z_2$ (Fig. 1c), and $\beta_{mp}(z,\lambda) = \beta_{mp}^{(a)}(\lambda)$ outside it. Substitution of expressions for the propagation constants from Eq. (1) into Eq. (2) makes Eq. (2) equivalent to the Schrödinger equation for the rectangular quantum well[32]. In particular, it becomes obvious that WGMs with wavelengths $\lambda_{mp}^{(a)} < \lambda < \lambda_{mp}^{(w)}$ can be localized by the droplet. These modes are oscillatory in the droplet region (here $\beta_{mp}(\lambda,z)$ is real) and exponentially decay outside it (here $\beta_{mp}(\lambda,z)$ is pure imaginary). The quantization rule for these modes is similar to the quantization rule for the rectangular quantum well. The transmission amplitude through the microfiber placed in contact with the capillary at point $z$ is expressed through the Green's function of Eq. (2), $G(z,z,\lambda)$ as[33]

$$A(z,\lambda) = A_0 - \frac{i|C|^2 G(z,z,\lambda)}{1 + DG(z,z,\lambda)}. \tag{3}$$

Here constants $A_0$, $|C|$, and $D$ are the microfiber–capillary coupling parameters, which are usually determined experimentally. As an example, Fig. 3c shows the surface plot of the transmission power $P(z,\lambda) = |A(z,\lambda)|^2$ for our microcapillary having $r_{int} = 61\,\mu m$ and $r_{ext} = 68\,\mu m$ with a water droplet of 100 μm length inside. The transmission power $P(z,\lambda)$ shown in Fig. 3c is found for the characteristic microfiber – capillary coupling parameters $A_0 = 0.8 - 0.1i$, $|C|^2 = 0.02\,\mu\text{m}^{-1}$ and $D = 0.02 + 0.025i\,\mu\text{m}^{-1}$ [33]. The cutoff wavelength of the empty capillary is set to $\lambda_{mp}^{(a)} = 1540\,\text{nm}$, close to one of the cutoff wavelengths observed in our experiment (Fig. 2a). Approximating the behavior of the $p = 3$ cutoff wavelength in Fig. 3a, we assume that this wavelength increases by $\Delta\lambda_{mp} = 1\,\text{nm}$ and becomes $\lambda_{mp}^{(w)} = 1.541\,\mu m$ in the droplet region. The introduced cutoff wavelength profile (bold black dashed line in Fig. 3c) corresponds to a quantum well in Eq. (2) and, equivalently, to an optical microresonator. The right hand side vertical axis in Fig. 3c is the effective radius variation determined from the cutoff wavelength variation by the rescaling equation $\Delta r_{mp}/r_{ext} = \Delta\lambda_{mp}/\lambda_{mp}^{(a)}$. The horizontal resonance pathways in Fig. 3c with interchanging nodes and antinodes correspond to the axial eigenvalues of this microresonator with quantum numbers $q = 0,1,...6$. In analogy to the bottle microresonators introduced by the external radius variation[25], this microresonator can be called a *droplet-induced bottle microresonator*.

### 4. Interpretation of the experimental observations

The surface plot of experimental spectra in Fig. 2a exhibit several series of resonance pathways. Some of pathways cross the droplet region without visible perturbation. As explained above, they correspond to the cutoff wavelengths with small radial quantum numbers $p < 2$. Other pathways experience a break in the droplet region which is accompanied by the appearance of series of resonance lines with interchanging nodes and antinodes. The behavior of each of these series, taken separately, is similar to that in the numerical example of a droplet-induced bottle resonator shown in Fig. 3c. To clarify this, we consider two examples. The first example corresponds to the behavior of the cutoff wavelength which is close to 1540.44 nm outside the droplet. This wavelength experiences the positive shift of 2.85 nm in the region of the droplet. The shift introduces a microresonator containing 15 eigenvalues with axial quantum numbers $q = 0,1,...,14$ indicated on the theoretical 2D plots of transmission power in Fig. 2b and 2c described below. Based on calculations shown in Fig. 3a, we suggest that the radial quantum number of these eigenvalues is equal to $p = 4$. As noted above, the process of characterization of the microresonator region took around 11 min. During this time the droplet became smaller in length due to evaporation. It is well known from quantum mechanics that the decrease of the length of the rectangular quantum well (resonator) leads to the growth of the separation between its eigenvalues[32]. For this reason, the pathways indicating the axial eigenvalues in Fig. 2a have a small negative slope. As shown in the next section, the measurement of this slope allows us to detect the variation of the droplet size with nanometer precision. Figs. 2b and 2c show the results of our theoretical characterization of the droplet using Eqs. (1)-(3) which fit the experimental spectra before and after measurements (see Supplementary note 1). The theoretical as well as experimental pathways with axial quantum numbers $q$ indicated in Fig. 2b consist of $q+1$ antinodes. The agreement between the positions of wavelength resonances found theoretically and their positions found experimentally is excellent. In the second example, we consider a 2D spectral fragment magnified in Fig. 2d. Here the cutoff wavelength equal to 1539.88 nm outside the droplet is shifted by 0.7 nm to 1540.58 nm in the droplet region. Based on calculations presented in Fig. 3a, we suggest that the radial quantum number of this cutoff wavelength is $p = 3$. Our numerical modelling shown in Fig. 2e and 2f precisely confirms the length of the droplet $L$ equal to 129.7 μm and 126.3 μm before and after its characterization, which has been initially calculated in the first example. Finally, Fig. 2g shows that the quality-factor of the introduced microresonator is remarkably large, $\sim 5 \cdot 10^5$.

Before the full stabilization of the droplet, it was translated along the originally dry capillary from right to left. As the result, tiny micro-islands of water adjacent to the internal capillary surface were left behind at the right hand side of the droplet. Examination of resonance pathways outside the droplet in Fig. 2a allows us to distinguish between the perturbations caused by outside contamination, which disturbs all pathways similarly, and perturbations caused by the water islands situated inside the capillary. As discussed above, the latter disturb the pathways with relatively large radial quantum numbers $p$ and have negligible effect on pathways with $p < 2$. For example, the pathway close to $\lambda = 1540.9$ nm, which crosses the droplet unperturbed and presumably has $p = 0$, is straight everywhere except of the disturbance near $z = 500$ μm. Since this disturbance is replicated in all other pathways, we suggest that it is caused by perturbations of the external capillary surface. However, the pathway close to $\lambda = 1540.5$ nm is straight at the left hand of the droplet and, unlike the pathway at $\lambda = 1540.9$ nm, is corrugated at its right hand side. For this reason, we suggest that these corrugations are caused by water islands inside the capillary.

### 5. Evaporation of the droplet and concomitant effects

Due to the evaporation of the droplet in the process of its characterization, the successive measurements comprising the 2D plot in Fig. 2a present the spectra of the microresonator continuously reducing in length. For this reason, in contrast to the purely horizontal resonance pathways in theoretical plots in Fig. 2b, c, e and f, the resonance pathways in the droplet region of experimental Fig. 2a are tilted. Analysis of the behavior of these tilted pathways allows us to determine the variation of the droplet length with exceptionally accurate nanometer precision. Fig. 2h shows the magnified fraction of the 2D plot in Fig. 2a, which includes the resonance pathway corresponding to axial quantum number $q = 12$ consisting of 13 antinodes. The most accurate measurement of the immediate wavelength eigenvalue of the microresonator is achieved at the nodes of this pathway, which correspond to the minimum coupling between the excited WGM and input-output microfiber. For the model of the rectangular quantum well described by Eqs. (1) and (2), the change $\Delta L$ of the resonator length $L$ can be determined from the shift $\Delta\lambda$ of the eigenvalue $\lambda_{mpq}$ with large axial quantum number $q \gg 1$ by the equation $\Delta L = 4n_{cap}^2 L^3 q^{-2} \lambda_{mp}^{-3} \Delta\lambda$ derived in Supplementary Note 2. From this equation and analysis of the pathway with axial quantum number $q = 12$ shown in Fig. 2h, we find that the total wavelength shift $\Delta\lambda_t = 0.11$ nm, which the eigenvalue $\lambda_{mp,12} \approx 1.541$ μm acquires during the time of scan of the droplet region equal to 10.7 min, corresponds to the droplet reduction in length of $\Delta L_t = 3.8$ μm and shrinking speed of 0.35 μm/min. Using the same equation, the resolution of the measurement of the droplet length variation is found from the resolution of our optical spectrum analyzer equal to $\Delta\lambda_r = 0.13$ pm as $\Delta L_r = 4.5$ nm. Due to geometric reasons, this remarkably small resolution is, nevertheless, two orders of magnitude larger than that achieved in SNAP technology for the effective variation of the optical fiber radius[3, 25, 27].

It is interesting to estimate the deformation of the capillary due to the surface tension introduced by the droplet. The additional pressure $P$ experienced by the capillary wall can be estimated by the Young-Laplace equation, $P = 2\gamma/R$ [34]. Assuming the characteristic value of the surface tension $\gamma \sim 0.1$ N/m and capillary and meniscus radii $r_{ext} \sim R \sim 100$ μm, we find $P \sim 1$ kPa. Depending on the actual values of the surface tension at the water-air, water-silica, and silica-air interfaces, which usually have the same order of magnitude[34], this pressure can be positive or negative. Consequently, it can increase or decrease the effective radius of the capillary. The radius variation $\delta r$ caused by the surface pressure can be found from the elasticity theory of thin-walled cylinders as $\delta r = Pr_{ext}^2 (E\Delta r)^{-1}$ where $\Delta r$ is the capillary wall thickness and $E$ is its Young modulus[35]. Since the effect of the stress-induced refractive index variation is usually smaller than that of the mechanical deformation, we estimate the corresponding shift of the capillary radius as $\delta r \sim 3$ pm and the cutoff wavelength as $\delta\lambda \sim \delta r \lambda / r_{ext} \sim \lambda_{mp} \gamma (E\Delta r)^{-1} \sim 0.1$ pm. These values are much smaller than those observed in our experiment for WGMs with radial quantum numbers $p > 2$. We suggest that the introduced shift can be detected by more accurate measurements of the cutoff wavelength shifts of WGMs with $p = 0$ since their evanescent coupling to the droplet is negligibly small.

Usually, a water droplet carries an electric charge $Q$ (see e.g., Refs.[36, 37]). The value of $Q$ is limited by the maximum charge-introduced expanding force which can be withheld by the surface tension. Estimating $Q$ by the Rayleigh formula for a spherical droplet, $Q \leq 8\pi(\varepsilon_0\gamma)^{1/2} R_0^{3/2}$, where $\varepsilon_0$ is the vacuum permittivity and $R_0 \sim 100$ μm is the characteristic radius of the droplet, we find $Q \leq 2\cdot 10^{-11}$ C. Assuming that the distribution of charge along the droplet surface is uniform we find that the charge-introduced effective radius variation of the capillary (limited by the droplet stability) has the same order of magnitude as the radius variation introduced by the surface tension calculated above. However, the electrostatic calculations (see e.g.[38]) show that the distribution of charge at the droplet-capillary surface interface can be strongly nonuniform. The charge density can grow by an order of magnitude near the droplet edges where the introduced capillary radius variation can be significantly greater.

Another effect introduced by the droplet charge is the DC Kerr effect[39, 40]. By modelling the droplet as a cylinder of length $L$ and charge $Q$, we estimate the electric field inside the capillary as $E = Q/(2\pi\varepsilon_0 n_{cap} L r_{ext})$. For $Q \sim 10^{-11}$ C, $L \sim 100$ μm and $r_{ext} \sim 50$ μm, we have $E \sim 10^7$ V/m. The refractive index variation introduced by this field is expressed through the third order electric susceptibility $\chi^{(3)}$ as $\delta n_{DCKerr} = (3\chi^{(3)}/2n_{cap})E^2$ [40]. For silica, $\chi^{(3)} \sim 2\cdot 10^{-22}$ m$^2/V^2$ and the refractive index variation of the capillary introduced by the DC Kerr effect in the droplet region is estimated as $\delta n_{DCKerr} \sim 10^{-7}$. This corresponds to the cutoff wavelength shift of $\delta\lambda \sim \delta n_{DCKerr} \lambda_{mp} / n_{cap} \sim 0.1$ pm having the same order of magnitude as the shifts introduced by the surface tension mechanically. From the above expressions for $E$ and $\delta n_{DCKerr}$ we have $\delta n_{DCKerr} \sim L^{-2}$. Therefore, the DC Kerr effect can be increased by an order of magnitude for a droplet with three times smaller length. In addition, the DC Kerr effect, being proportional to the local charge density squared, is significantly greater near the droplet edges where, as noted above, the charge density maximizes.

## 6. Discussion

We have shown, experimentally and theoretically, that a droplet positioned inside an optical microcapillary can trap the WGMs propagating along the capillary wall and localize them completely. The evanescent tails of WGMs penetrate into a thin layer of the droplet through the internal capillary surface. Consequently, the WGMs spectra characterize the medium which is adjacent to the interface between the droplet and the capillary. In contrast to previous WGM microfluidic sensing methods[13-16], the presence of the droplet-introduced microresonator allows us to monitor the droplet dynamics nonlocally with the fixed position of the input-output microfiber. For example, the resonance spacing measured at a single point $z_0$ in the droplet region (Fig. 1c) allowed us to determine the immediate length of the droplet which changed in the process of evaporation. More generally, examination of the WGM spectra measured at one or a few points along the droplet enables monitoring the behavior of the droplet away from these points. The important question remained is whether the collected spectral data is sufficient to solve the inverse problem completely and determine the refractive index variation near the internal capillary surface.

The future research and development exploring the discovered phenomenon promises wide-ranging applications in microfluidics, surface science, and microphotonics. It is of great interest to apply the developed method to investigate the droplets of different liquids, simple and complex, including colloidal liquids[34]. As an example, the immediate positions and velocities of an individual and, presumably, a few micro/nanoparticles in a colloidal liquid can be determined from the dynamics of WGM spectrum measured at a fixed microfiber position. In contrast to the SNAP microfluidic sensor[28, 29], the microresonator enabling this nonlocal characterization of the droplet is naturally introduced and does not have to be fabricated.

It is expected that the size of particles which can be detected and monitored with our approach can achieve the sub-nanoscale dimensions. In fact, it has been shown that a WGM microsphere resonator sensor enriched with plasmonic nanoparticles can detect single atoms [15]. Due to the analogy between the behavior of evanescent WGM at the periphery of a microsphere and at the interior of a capillary, we suggest that the similar resolution is feasible for plasmon-enhanced WGM sensing inside a microcapillary. Remarkably, our method potentially enables the detection of the immediate axial coordinate of moving nanoparticles and molecules, which was not possible in the previous approaches. More complex phenomena at the solid-liquid interfaces can be investigated[41-44].

**Acknowledgements**

The authors acknowledge funding from the Horizon 2020 Framework Programme (H2020) (H2020-EU.1.3.3, 691011) and Engineering and Physical Sciences Research Council (EPSRC) (EP/P006183/1). M.S. acknowledges the Royal Society Wolfson Research Merit Award (WM130110).

**Author contributions**

T.H. and K.V.T. performed the experiment and participated in data analysis. B.J.M. and R.S.W. fabricated the microcapillary fiber for the experiment. A.V.D. participated in the experiment and data analysis and co-supervised the project at its initial stage. D.L.P.V. participated in the experimental work and data analysis. P.B. co-supervised the project and participated in data analysis. M.S. conceived the experiment, supervised the project, developed the theory of the observed phenomenon, and wrote the manuscript. All authors participated in the discussion of results and final editing of the manuscript.

**Methods**

**Fabrication of the microcapillary**
The capillary fiber was drawn at OFS Laboratories from a tube-shaped preform fabricated of pure silica. The wall thickness of the capillary was controlled by the original dimensions of the tube and the internal pressure introduced in the process of drawing. The capillary fiber was coated with the protective polymer. In the experiment, the polymer coating was removed in hot sulfuric acid and the extracted silica capillary was cleaned in deionized water.

**Fabrication of the input-output fiber taper**
The input-output fiber taper with micron-diameter waist was fabricated in the sapphire microfurnace described in Ref. [45]. The commercial single mode fiber was stripped from the polymer coating and aligned inside the sapphire capillary heated with the $CO_2$ laser. The tapering method, which included pulling the fiber with two linear stages, was similar to the flame-brushing method developed for the fabrication of biconical tapers softened in flame.

**Measurement of the transmission power**
The measurement of the WGM spectrum was performed using the input-output fiber taper having the micron-diameter waist (microfiber) oriented normally to the capillary. The microfiber was translated along the capillary at a distance of 0.5 mm and periodically, every 2 μm, was placed in direct contact with the capillary. The transmission power was measured at each contact point by the Luna Optical Spectrum Analyzer with the wavelength resolution 1.3 pm. The measurement process was automated with the AutoHotkey software, which allowed us to periodically switch between the spectral measurement at contact points and translation of the microfiber along the capillary. The results of measurements were collected in the surface plot of transmission amplitude $P(z,\lambda)$ shown in Fig. 2a above and in Fig. S1.1 in Supplementary note 1.

**Modelling of the microresonator**
For the rectangular dependence of the propagation constant considered, the Green's function of Eq. (2) can be determined analytically. Consequently, the analytical expression for the transmission amplitude, Eq. (3), can be found. Nevertheless, the theoretical plots of transmission power shown in Fig. 2b, c, e, f and 3c were obtained by numerical solution of Eq. (2) based on the Mathcad code which was written and verified previously[3, 25, 33]. In the numerical analysis of the experimental data, we varied the width and height of the rectangular resonator described by Eq. (1)-(3) to fit the spacing of the axial resonances at points $z_1$ and $z_2$ (Fig. 1c), i.e., immediately before (after) the input-output microfiber enter (exit) the droplet region. This analysis including is described in Supplementary Note 1.

**Data availability**
Contact the corresponding author for the additional data supporting the results of this study.

## Supplementary information

**Supplementary note 1**

**1.1 Expanded experimental data and processing details**

The 2D plot of the transmission power $P(z, \lambda)$ shown in Fig. 2a of the main text is a fragment of the 2D plot measured in an expanded 10 nm bandwidth shown in Fig. S1.1. The spectral quasiperiod equal to 3.92 nm was measured from Fig. S1.1. The black rectangle in this figure, which contains a single period, outlines the region shown in Fig. 2a.

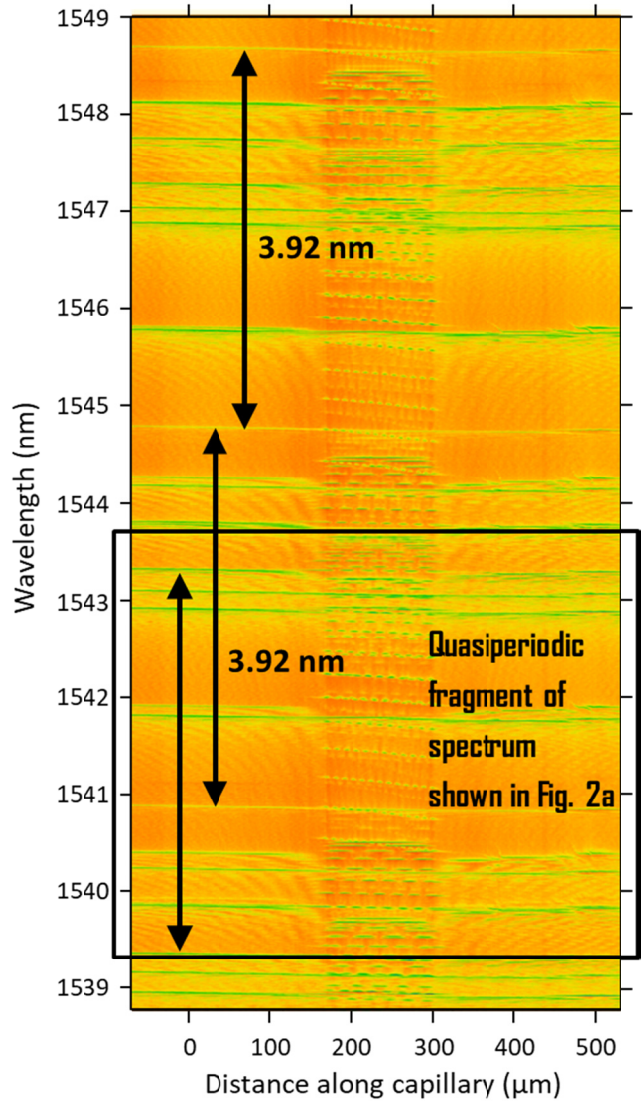

**Fig. S1.1.** The surface plot of the transmission power spectra in the bandwidth from 1539 nm to 1549 nm. The spectral (vertical axis) and spatial (horizontal axis) resolutions of the scan were 1.3 pm and 2 μm, respectively.

Fig. S1.2 illustrates the method we used to analyze the experimental data and, in particular, to determine the initial and finite droplet widths, the shift of the cutoff wavelength due to the presence of water, and the speed of the droplet evaporation. The left hand side (theory) and right hand side (experiment) plots in this figure are fragments taken from the theoretical and experimental plots of Fig. 2.

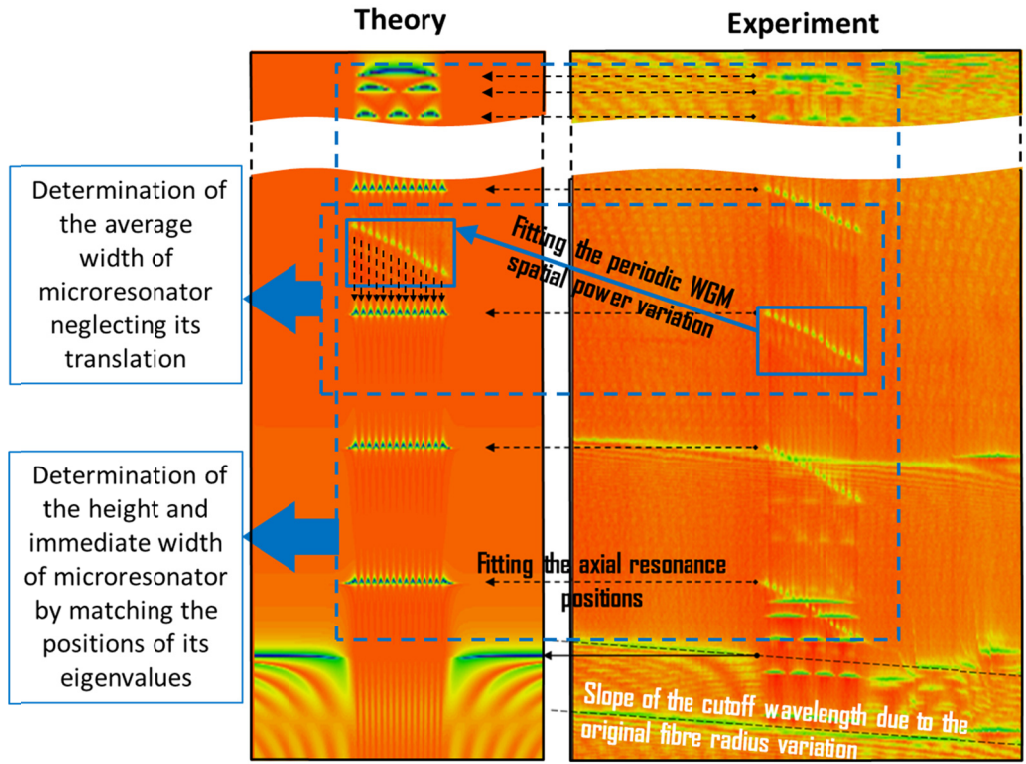

**Fig. S1.2.** Illustration of matching the theoretical and experimental data. The width of the resonator is determined by the comparison of single resonance pathways. The height of the resonator is determined by fitting the positions of all eigenvalues corresponding to the selected cutoff wavelength.

With a good accuracy, the axial coordinates $z_1$ and $z_2$ of the contact circumferences between the droplet meniscuses and the capillary (Fig. 1c) correspond to the actual edges of the droplet-introduced resonator. In fact, since the contact angle between the meniscuses and capillary is finite, the characteristic transition length between the cutoff wavelength of the empty and water-filled capillary has the same sub-micron scale as the penetration depth of the WGMs into the air. The latter fact justifies the model of rectangular resonator used in our simulations.

The experimental plot in Fig. S1.2 consists of magnified fragments of the experimental theoretical plots shown in Fig. 2a and 2b, respectively. The measured width of the droplet was first determined directly from the experimental Fig. 2a. Next, in a more accurate measurement, this width was determined from the comparison of numerical calculations of the transmission spectra using Eqs. (1)-(3) with the experimental data in Fig. 2a. The comparison method is illustrated in Fig. S1.2. We translated the square fragment of the experimental plot with an eigenvalue pathway inside outlined by a blue solid rectangular onto the theoretical plot of Fig. S1.2. The width of the resonator was determined by fitting the positions of the nodes of the experimental and theoretical plot indicated by vertical arrows. This method determines the final width of the resonator and does take into account its possible small displacement in the process of measurement.

Using the determined value of the resonator width as a first approximation, the cutoff wavelength shift (the resonator height) and immediate initial and finite resonator widths were found by matching the theoretical and experimental positions of the resonator eigenvalues. This was performed for all the eigenvalues corresponding to the selected cutoff wavelength. The process of matching included both the variation of the resonator height and small adjustment of the resonator width. The remarkably accurate matching of these positions illustrated in Fig. S1.2 justified the rectangular model of the resonator. Notice that the values of the cutoff wavelengths shifts are independent of the droplet size and do not change in the process of measurement. For the examples considered in Fig. 2, these values, equal to 3.85 nm and 0.7 nm, were first determined experimentally and then adjusted numerically in the process of fitting the wavelength eigenvalue positions. This fact allowed us to determine the actual initial and finite widths of the resonator and exclude the effect of its small translation.

# Supplementary note 2

## 2.1 Semiclassical quantization rule for WGMs in the droplet-introduced microresonator

If the WGM coupling to the capillary interior is ignored, the cutoff wavelengths $\lambda_{mp}^{\pm}$ can be analytically determined for large azimuthal quantum numbers $m \gg p$ by the asymptotic equation (see e.g., Ref. [1]):

$$\lambda_{mp}^{\pm} \approx \frac{2\pi n_{cap} r_{ext}}{m}\left(1 + \varsigma_p\left(2m^2\right)^{-1/3} + \frac{n_{cap}^{\pm 1}}{m(n_{cap}^2 - 1)^{1/2}}\right) \quad (S2.1)$$

where $\pm$ corresponds to the TE/TM polarization and $\varsigma_p$ are the zeros of the Airy function, $\varsigma_0 = 2.338$, $\varsigma_1 = 4.088$ $\varsigma_2 = 5.521$, $\varsigma_3 = 6.787$, and $\varsigma_p \approx (\frac{3}{8}\pi(4p-1))^{2/3}$ for $p \gg 1$. From Eq. (S2.1), at wavelength around $\lambda_0 = 1.54$ µm the spectrum of cutoff wavelengths of our microcapillary is quasiperiodic along the azimuthal quantum number with the period of $\Delta\lambda = 3.95$ nm. The latter value is in excellent agreement with the experimental $\Delta\lambda = 3.92$ nm found from Fig. S1. The cutoff wavelengths plotted in Fig. 3b are found from Eq. (S2.1).

The quantization rule for the WGMs localized in the region of the droplet $z_1 < z < z_2$ (Fig. 1c) is written as

$$\beta_{mp}^{(w)}(\lambda_{mpq}^{(w)})L = \pi(q + \kappa), \qquad q = 0,1,2,... \quad (S2.2)$$

where $L = z_2 - z_1$ is the droplet length and $\kappa \sim 1$ is the parameter determined by matching solutions of Eq. (2) at the droplet boundary[2]. Substitution of the expression for $\beta_{mp}^{(w)}(\lambda)$ given by Eq. (1) into Eq.(S2.2) yields the wavelength eigenvalues of the droplet introduced microresonator:

$$\lambda_{mpq}^{(w)\pm} = \lambda_{mp}^{(w)\pm} - \frac{(q+\kappa)^2}{8n_{cap}^2 L^2}\left(\lambda_{mp}^{(w)\pm}\right)^3 \quad (S2.3)$$

For large $q \gg 1$ we can ignore $\kappa$ in Eq. (S2.3) and the variation $\Delta L$ of microresonator length $L$ is expressed through the variation $\Delta\lambda$ of eigenvalue $\lambda_{mpq}^{(w)\pm}$ by the equation

$$\Delta L = \frac{4n_{cap}^2 L^3}{q^2 \left(\lambda_{mp}^{(w)\pm}\right)^3}\Delta\lambda \quad (S2.4)$$

used in the main text.

## 2.2 Derivation of Eq. (1) for the propagation constant of slow WGMs

The expression for the propagation constant of a slow WGM given by Eq. (1) can be derived using the first order perturbation theory[2]. The solution of Maxwell's equations in the axially symmetric optical fiber (capillary) in the cylindrical coordinates $(\rho, z, \varphi)$ as $\exp(\pm i\beta_{mp}(\lambda)z)\exp(im\varphi)Q_{mp}(\rho, \lambda)$ where function $Q_{mp}(\rho, \lambda)$ satisfies the differential equation[3]:

$$\frac{d^2 Q_{mp}(\rho,\lambda)}{d\rho^2} + \frac{1}{\rho}\frac{dQ_{mp}(\rho,\lambda)}{d\rho} + \left[\left(\frac{2\pi n(\rho)}{\lambda}\right)^2 - \beta_{mp}^2(\lambda) - \frac{m^2}{\rho^2}\right]Q_{mp}(\rho,\lambda) = 0, \quad (S2.5)$$

where $n(\rho)$ is the radial distribution of the refractive index and integers $m$ and $p$ are the azimuthal and radial quantum numbers. For WGMs having $m \gg 1$, the cutoff frequencies $\lambda_{mp}$ correspond to the vanishing propagation constant:

$$\beta_{mp}(\lambda_{mp}) = 0 \quad (S2.6)$$

We rewrite Eq. (S2.5) as:

$$\mathbf{L} Q_{mp}(\rho,\lambda) = \beta_{mp}^2(\lambda) Q_{mp}(\rho,\lambda)$$
$$\mathbf{L} = \frac{d^2}{d\rho^2} + \frac{1}{\rho}\frac{d}{d\rho} + \left(\frac{2\pi n(\rho)}{\lambda}\right)^2 - \frac{m^2}{\rho^2} \quad \text{(S2.7)}$$

It is seen from Eq. (S2.7) that $\beta_{mp}^2(\lambda)$ is the eigenvalue of the operator $\mathbf{L}$. Consequently, from Eq. (S2.6), functions $Q_{mp}(\rho,\lambda_{mp})$ are the eigenfunctions of the operator

$$\mathbf{L}_{mp}^{(0)} = \frac{d^2}{d\rho^2} + \frac{1}{\rho}\frac{d}{d\rho} + \left(\frac{2\pi n(\rho)}{\lambda_{mp}}\right)^2 - \frac{m^2}{\rho^2} \quad \text{(S2.8)}$$

and have zero eigenvalues. For small $\beta_{mp}(\lambda)$ corresponding to small deviation of $\lambda$ from $\lambda_{mp}$ we have:

$$\mathbf{L} = \mathbf{L}_{mp}^{(0)} + \Delta \mathbf{L}_{mp}^{(0)},$$
$$\Delta \mathbf{L}_{mp}^{(0)} = \left(\frac{2\pi n(\rho)}{\lambda}\right)^2 - \left(\frac{2\pi n(\rho)}{\lambda_{mp}}\right)^2 \approx \frac{2(2\pi n(\rho))^2}{\lambda_{mp}^3}(\lambda_{mp} - \lambda), \quad \text{(S2.9)}$$

Using Eq. (S2.9), we find eigenvalues $\beta_{mp}^2(\lambda)$ in the first order of the perturbation theory[2]:

$$\beta_{mp}^2(\lambda) = \langle Q_{mp}(\rho,\lambda_{mp}) | \Delta \mathbf{L}_{mp}^{(0)} | Q_{mp}(\rho,\lambda_{mp}) \rangle = \frac{8\pi^2}{\lambda_{mp}^3} \langle Q_{mp}(\rho,\lambda_{mp}) | n^2(\rho) | Q_{mp}(\rho,\lambda_{mp}) \rangle (\lambda_{mp} - \lambda) \quad \text{(S2.10)}$$

where functions $Q_{mp}(\rho,\lambda_{mp})$ are normalized, $\langle Q_{mp}(\rho,\lambda_{mp}) | Q_{mp}(\rho,\lambda_{mp}) \rangle = 1$. Since $Q_{mp}(\rho,\lambda_{mp})$ is localized in the region of the capillary wall where $n(\rho) = n_{cap} = const$, we can ignore the contribution of evanescent parts of $Q_{mp}(\rho,\lambda_{mp})$ in Eq. (S2.10). Then Eq. (S2.10) is simplified to

$$\beta_{mp}^2(\lambda) = \frac{8\pi^2}{\lambda_{mp}^3} n_{cap}^2 (\lambda_{mp} - \lambda). \quad \text{(S2.11)}$$

This equation is identical to Eq. (1) of the main text.

**References**

1. Demchenko YA and Gorodetsky ML. Analytical estimates of eigenfrequencies, dispersion, and field distribution in whispering gallery resonators. J. Opt. Soc. Am. B 2013; 30: 3056-3063.
2. Landau LD and Lifshitz EM. Quantum Mechanics. Pergamon Press, 1965.
3. Snyder AW and Love J, Optical waveguide theory. Chapman and Hall, 1983.